\documentclass[twocolumn]{aastex631}

\begin{document} 
\title{Primordial Black Hole-Neutron Star Merger Rate in Modified Gravity}

\author[0000-0002-6349-8489]{Saeed Fakhry} 
\email{s\_fakhry@sbu.ac.ir}
\affiliation{Department of Physics, Shahid Beheshti University, 1983969411, Tehran, Iran}

\author{Maryam Shiravand}
\email{ma\_shiravand@sbu.ac.ir}
\affiliation{Department of Physics, Shahid Beheshti University, 1983969411, Tehran, Iran}

\author{Marzieh Farhang}
\email{m\_farhang@sbu.ac.ir}
\affiliation{Department of Physics, Shahid Beheshti University, 1983969411, Tehran, Iran}

\begin{abstract}
\noindent
In this work we investigate the merger rate of primordial black hole-neutron star (PBH-NS) binaries in two widely-studied modified gravity (MG) models: Hu-Sawicki $f(R)$ gravity and the normal branch of Dvali-Gabadadze-Porrati (nDGP) gravity. In our analysis, we take into account the effects of MG on the halo properties including: halo mass function, halo concentration parameter, halo density profile, and velocity dispersion of dark matter particles. We find that these MG models, due to their stronger gravitational field induced by an effective fifth force, predict enhanced merger rates compared to general relativity. This enhancement is found to be redshift-dependent and sensitive to model parameters, PBH mass and fraction. Assuming PBH mass range of $5\mbox{-}50\,M_\odot$, we compare the predicted merger rate of PBH-NS binaries with those inferred from LIGO-Virgo-KAGRA observations of gravitational waves (GWs). We find that the merger rates obtained from MG models will be consistent with the GW observations, if the abundance of PBHs is relatively large, with the exact amount depending on the MG model and its parameter values, as well as PBH mass. We also establish upper limits on the abundance of PBHs in these MG frameworks while comparing with the existing non-GW constraints, which can potentially impose even more stringent constraints.
\end{abstract}


%%%%%%%%%%%%%%%%%%%%%%%%%%%%%%%%%%%%%%%%%%%%%%%%
\section{Introduction} 
For many years, gravitational waves (GWs) have been a focal point in the field of cosmology, offering a novel approach to explore a variety of astrophysical and cosmological events. The merger of compact binary systems is considered as a potential source of GW emission \citep{2022LRR....25....1M}. In recent years, GW detectors have documented numerous events resulting from compact binary mergers \citep[e.g.,][]{2016PhRvL.116f1102A, 2016PhRvL.116x1103A, 2016PhRvL.116v1101A, 2020ApJ...896L..44A, 2020PhRvL.125j1102A}. Compact binaries are typically classified into three categories: binary black holes (BBH), black hole-neutron star (BH-NS) binaries, and binary neutron stars (BNS). Among these, most recorded GWs are attributed to BBH mergers in the mass range of $10\mbox{-}100\,M_{\odot}$ \citep[e.g.,][]{2019PhRvX...9c1040A, 2021PhRvX..11b1053A, 2023PhRvX..13d1039A}. While the genesis of blach holes (BHs) has been thoroughly investigated, the precise process of their formation remains an active area of study \citep[e.g.,][]{2019JCAP...02..018R, 2021MNRAS.507.5224B, 2021PhRvL.126b1103N, 2021PhRvL.127o1101N, 2022PhR...955....1M, 2023ApJ...952..121L}. They might have originated from stellar collapse through various channels \citep{2021ApJ...912...98F, 2021RNAAS...5...19R}, resulting in what is known as astrophysical black holes (ABHs). Alternatively, they could be remnants from the early Universe, where density fluctuation-induced gravitational collapse could have led to the formation of primordial black holes (PBHs) \citep{1967SvA....10..602Z, 1971MNRAS.152...75H}. 

PBHs, with their diverse masses \citep[e.g.,][]{2018CQGra..35f3001S}, are thought to stand out from ABHs and are considered potential dark matter candidates \citep[for other possible candidates, see, e.g.,][]{2000PhRvL..84.3760S, 2000ApJ...542..281A, 2012LRR....15....6L, 2018RPPh...81f6201R, 2020PhRvD.102h4063D, 2021CQGra..38s4001D}. Over the years, robust observational techniques have been used to constrain the abundance of PBHs across various mass scales. These observations have provided reliable frameworks for studying the early Universe at small scales \citep[e.g.,][]{2017PhRvD..96b3514C, 2018JCAP...04..007L, 2021RPPh...84k6902C}. Furthermore, the presumption of their involvement in merger events linked to GW detectors allows for the acquisition of robust limitations on the abundance of PBHs. In spite of numerous theoretical ambiguities, PBHs of stellar mass (those in line with GW observations) might play a substantial role in the composition of dark matter \citep{2016PhRvL.116t1301B, 2017PDU....15..142C}. Nevertheless, the hypothesis of binary PBH genesis in the early Universe implies that the contribution of such Black Holes to the structure of dark matter ought to be minimal to align with the observations of the LIGO-Virgo-KAGRA (LVK) detectors \citep{2016PhRvL.117f1101S}.

Moreover, the LVK detectors are able to identify a distinct kind of GW event referred to as BNS merger events \citep{2020ApJ...888L..10Y}. BNS systems can dynamically originate in high-density areas like star clusters. These events emit not only GWs but also electromagnetic waves, rendering them an important topic for multi-messenger astronomy research. It is important to highlight that distinguishing between NS candidates and solar-mass BHs poses a challenge without separate analysis of electromagnetic signals and setting suitable upper bounds on the tidal deformability parameter of BNS \citep{2017PhRvL.119p1101A, 2017ApJ...848L..12A}. This issue has been tackled in prior research, such as the studies conducted by \cite{2017PhRvL.119p1101A, 2017ApJ...848L..12A}. Currently, the field of multi-messenger astronomy, which explores this problem, is leading the research in this domain \citep[e.g.,][]{2021JCAP...10..019T, 2021PhRvL.126n1105D}.

Furthermore, BH-NS binary mergers can make a significant contribution to the data collected by the LVK detectors \citep{2021FrASS...8...39R}. Such events can yield valuable insights for multi-messenger observations and can generate both GWs and electromagnetic signals during the merger phase \citep{2020EPJA...56....8B}. Typically, these events involve a post-merger phase where the residual matter from the NS is accreted by the BH, leading to a luminous event \citep{2021ApJ...923L...2F}. BH-NS binary mergers are of particular interest as they provide unique perspectives into the nuclear equation of state of NSs and the accretion processes of BHs \citep[see, e.g.,][]{2019PhRvD.100f3021H, 2019PhRvL.123d1102Z, 2021ApJ...918L..38F, 2021PhRvD.104l3024T}. Moreover, they can aid in constraining the spin and abundance of BHs. Recently, the LVK detectors reported the first two direct events related to BH-NS mergers. The estimated mass components of these events were ($8.9^{+1.2}_{-1.5}\,M_{\odot}, 1.9^{+0.3}_{-0.2}\,M_{\odot}$) and ($5.7^{+1.8}_{-2.1}\,M_{\odot}, 1.5^{+0.7}_{-0.3}\,M_{\odot}$) \citep{2021ApJ...915L...5A}.

The formation process of BH-NS binaries is still fraught with many uncertainties. As previously mentioned, one possible assumption is that the contributing BHs are of primordial origin. However, unlike the formation of PBH binaries, the formation of PBH-NS binaries is associated with the late-time Universe, taking place immediately after the formation of cosmological and astrophysical structures. As a result, it is anticipated that more precise models describing dark matter halos will significantly influence the density and velocity distribution of PBHs involved in these events \citep[see, e.g.,][]{2021PhRvD.103l3014F, 2022PhRvD.105d3525F, 2022ApJ...941...36F, 2023PDU....4101244F, 2023PhRvD.107f3507F, 2023ApJ...947...46F, 2023arXiv230811049F}. Various methods have been utilized to elaborate on halo characteristics within the context of GR \citep{1974ApJ...187..425P, 1999MNRAS.308..119S, 2001MNRAS.323....1S, 2001MNRAS.321..372J, 2003MNRAS.346..565R, 2006ApJ...646..881W, 2007MNRAS.374....2R, 2017JCAP...03..032D, 2023PDU....4101259D, 2023arXiv231115307F}. However, a critical question arises as to whether the principles of GR and their derived concepts adequately explain the formation and evolution of large-scale structures such as galactic halos. Currently, there is growing consensus that while GR has been successful in accurately predicting numerous events, it falls short in fully elucidating large-scale phenomena, especially within the dark sectors of the Universe \citep[for more details on see, e.g.,][]{2014JCAP...08..034Y, 2018RPPh...81a6901H, 2021arXiv210700562G, 2023PDU....3901144G}. Consequently, considerable efforts have been devoted to generalizing the Einstein-Hilbert action, known as modified gravity (MG) \citep[see. e.g.,][]{2007IJGMM..04..115N, 2012PhR...513....1C, 2021arXiv210512582S}.

To fulfill the observational criteria related to gravity, such as restoring GR in extensively tested high-density scenarios and aligning with the expansion history \citep{2019LRR....22....1I}, MG theories necessitate additional screening mechanisms \citep[e.g.,][]{2018arXiv181002652A}. The Chameleon and Vainshtein effects are two commonly studied mechanisms \citep{1972PhLB...39..393V, 2013ApJ...779...39J}. The operation of these screening mechanisms is the most influential factor distinguishing different MG models, and their interaction and response with the Large Scale Structure and the cosmological environment play a pivotal role in governing the effects induced by MG theories. In this study, we limit our investigation to two widely recognized MG models that adhere to the screening mechanism. The first model, referred to as Hu-Sawicki $f(R)$ gravity, incorporates additional scalar fields and their interaction with matter, allowing for nonlinear functions of the Ricci scalar \citep{2007PhRvD..76f4004H}. The second model, known as the normal branch of the Dvali-Gabadadze-Porrati (nDGP) model, explores the possibility that gravity propagates in extra dimensions, distinguishing it from other conventional forces \citep{2000PhLB..485..208D}. Both of these complex MG theories share a common characteristic: the emergence of a fifth force on cosmological scales, which arises due to the existence of extra degrees of freedom. In such theories that modify GR on large scales, the evolution of perturbations deviates from what is predicted by the standard model of cosmology. This modification introduces a fifth force that is expected to influence the formation of structures in the Universe \citep[see, e.g.,][]{2008PhRvD..77b4024C, 2016MNRAS.459.3880H, 2020PhRvD.102d3501C, 2021MNRAS.508.4140M}. A notable feature of the extra degrees of freedom in the aforementioned MG models is their adaptability depending on the environment, which is known as their chameleon nature. Specifically, they are light in areas of low density but heavy in high-density regions of matter \citep{2013JCAP...04..029B, 2016ARNPS..66...95J, 2018LRR....21....1B}.

In light of these discussions, it is plausible to anticipate that MG theories may result in alterations to the theoretical constructs of dark matter halos. Hence, the objective of this research is to calculate the merger rate of PBH-NS binaries within the context of Hu-Sawicki $f(R)$ and nDGP gravitational models. In this regard, the structure of the study is organized as follows: Section \ref{sec:ii} is dedicated to outlining the theoretical underpinnings of MG models. In Section \ref{sec:iii}, we explain the principles of dark matter halo models and key elements such as the halo density profile, the halo concentration parameter, and the halo mass function, all within the theoretical frameworks of MG models. Furthermore, in Section \ref{sec:iv}, we delineate the merger frequency of PBH-NS binaries in MG models and compare it with the analogous results from GR. In Section \ref{sec:v}, we deliberate on the results. Lastly, in Section \ref{sec:vi}, we encapsulate the findings and draw conclusions.
%%%%%%%%%%%%%%%%%%%%%%%%%%%%%%%%%%%%%%%%%%%%%%%%
\section{Modified Gravity Models} \label{sec:ii} 
%%%%%%%%%%%%%%%%%%%%%%%%%%%%%%%%%%%%%%%%%%%%%%%%
In this section, we focus on two fundamental models discussed in the field of MG literature: the Hu-Sawicki $f(R)$ model and the nDGP model. These models embody the screening mechanisms known as the Chameleon and Vainshtein classe.
\subsection{Hu-Sawicki $f(R)$ Model}
The Hu-Sawicki $f(R)$ model involves introducing a non-linear modification function, denoted as $f(R)$, to the conventional Einstein-Hilbert action \citep{2007PhRvD..76f4004H}
\begin{equation}\label{frgravity}
S=\int d^{4}x \sqrt{-g}\left[\frac{R+f(R)}{2\kappa}+\mathcal{L}_{\rm m}\right],
\end{equation}
where $R$ is the Ricci scalar, $\kappa$ is the Einstein gravitational constant, $g$ is the metric determinant, and $\mathcal{L}_{\rm m}$ is the matter Lagrangian. By selecting $f = -2\Lambda$, it is worth noting that one can restore GR while also incorporating a cosmological constant.

By employing a conformal transformation, Eq.\,(\ref{frgravity}) can be converted to a scalar-tensor theory involving the scalaron, denoted as $f_{\rm R} \equiv {\rm d}f(R)/{\rm d}R$. This scalaron serves as the degree of freedom introduced by MG. Selecting an appropriate functional form for $f(R)$ is crucial to ensure the occurrence of cosmic acceleration in the late-time Universe while still adhering to the limitations imposed by the solar system tests \citep{2008PhRvD..78j4021B}. A class of broken power-law models, referred to as the Hu-Sawicki model, is proposed in \cite{2007PhRvD..76f4004H} that effectively addresses the aforementioned situation. Specifically, this model can be described as follows
\begin{equation}
f_{R}=-m^2\frac{c_1(R/m^2)^n}{c_2^2(R/m^2)^n+1},
\end{equation}
where $m^2=\kappa\bar{\rho}_{m0}/3$ represents the characteristic mass scale, and $\bar{\rho}_{m0}$ denotes the background density of matter in the present-time Universe. Also, $c_1$, $c_2$, and $n>0$ are dimensionless free parameters that need to be determined in a specific manner to accurately recover the expansion history and pass solar-system tests, using the chameleon mechanism.

Note that the stability of the solution in high-density areas, i.e., where $R \gg m^2$, has to be maintained. In addition, cosmological experiments based on the $f(R)$ model should align with experiments based on GR. To fulfill this requirement, one must have $f_{RR}={\rm d}^2 f/{\rm d}R^{2} > 0$. Hence, one can expand the Hu-Sawicki model as follows
\begin{equation}\label{limfr}
\lim_{m^2/R \rightarrow 0} f(R) \approx -\frac{c_1}{c_2}m^2+\frac{c_1}{c_2^2}m^2\left(\frac{m^2}{R}\right)^n.
\end{equation}

Despite the absence of an actual cosmological constant in the Hu-Sawicki model, the model exhibits characteristics resembling a cosmological constant in both large-scale and local experiments. Moreover, the finite value of $c_1/c_2$ results in a constant curvature that remains unaffected by variations in matter density. Consequently, this choice allows for a class of models capable of accelerating the expansion of the Universe, which is similar to the behavior observed in the standard model of cosmology. Therefore, Eq.\,(\ref{limfr}) can be reformulated as
\begin{equation}
f(R) = -\frac{c_1}{c_2}m^2-\frac{f_{R_{0}}}{n}\frac{\bar{R}_{0}^{n+1}}{R^n},
\end{equation}
where $\bar{R}_{0}$ denotes the present-day background curvature, and $f_{R0} \equiv f_R(\bar{R}_0)$ is the field strength \footnote{If $|f_{R0}|\rightarrow 0$, one can obtain $(c_1/c_2) m^2 = 2\kappa\bar{\rho}_{\Lambda}$, where $\bar{\rho}_{\Lambda}$ represents the inferred background energy density attributed to dark energy.}. Cosmological and solar-system tests have limited the field strength $f_{R0}$ \citep{2020PhRvD.102j4060D}. In this regard, various values of $f_{R0}$ have been explored in the literature, spanning from $10^{-4}$ to $10^{-8}$ \citep{2019JCAP...09..066M}. In our study, we concentrate on the Hu-Sawicki model with $n = 1$ and $|f_{R0}| = 10^{-4}, 10^{-5}$, and $10^{-6}$, denoted as f4, f5, and f6, respectively.

%----------------------------------------
\subsection{nDGP Model}
The normal branch Dvali-Gabadadze-Porrati (nDGP) model of gravity is a modified gravity theory proposed in \cite{2000PhLB..485..208D}. The nDGP model describes the Universe as a four-dimensional brane embedded in a five-dimensional Minkowski space. The model assumes an action comprising two terms as
\begin{equation}\label{frgravity}
S=\int d^{4}x \sqrt{-g}\left[\frac{R}{2\kappa}+\mathcal{L}_{\rm m}\right] + \int d^{5}x \sqrt{-g_5}\frac{R_{5}}{2\kappa_{5} r_{\rm c}},
\end{equation}
where $R_5$, $g_5$, and $\kappa_5$ represent the Ricci scalar, metric determinant, and Einstein gravitational constant of the fifth dimension, respectively. Moreover, $r_{\rm c}=(\kappa_5/2 \kappa)$ denotes the crossover distance, which signifies a specific scale below which GR transforms into a four-dimensional model. For scales bigger than $r_{\rm c}$, the second term in the action becomes dominant, resulting in expected deviations from GR.

This model is composed of two branches: the ``normal" branch (nDGP), provided here, and the self-accelerating branch wich is called sDGP. We concentrate on the former since it is free of ghost instabilities \citep[e.g.,][]{2009PhRvD..80f3536L}. At larger scales, gravity becomes stronger, while at smaller scales, gravity behaves like GR due to Vainshstein screening. Therefore, one can investigate the nDGP model coupled with dark energy to match the desired $\Lambda$CDM expansion history. This path remains compelling because of prior investments in simulations. In this case, the only adjustable parameter to be constrained is $n = H_0r_{\rm c}$ (where $H_0$ stands for the Hubble constant), with values between $1$ and $5$ being extensively studied. Note that GR will be recovered if $n\rightarrow \infty$, which corresponds to a large gradients of gravitational force in Vainshtein screening. The nDGP model has been the subject of various studies and simulations to understand its implications for structure formation and cosmology, and its properties were investigated through numerical simulations and comparisons with observational data \citep[see, e.g.,][]{2021MNRAS.503.3867H, 2022JCAP...12..028F, 2023PhRvD.107d3533N}. In this work, we adopt three permissible values of $n=1$, $2$, and $5$, and we denote the models associated with these values as nDGP(1), nDGP(2), and nDGP(5), respectively.
%%%%%%%%%%%%%%%%%%%%%%%%%%%%%%%%%%%%%%%%%%%%%%%%%%%%%%%%%%%%%%%%
\section{Dark Matter Halo Models}\label{sec:iii}
%%%%%%%%%%%%%%%%%%%%%%%%%%%%%%%%%%%%%%%%%%%%%%%%%%%%%%%%%%%%%%%%
In this section, we provide a brief overview of dark matter halo models derived from MG models. These models are used as initial inputs in calculating the merger rate of binary systems consisting of PBHs and NSs.
\subsection{Halo Density Profile}
Within the framework of cosmological perturbation theory, dark matter halos are identified as dynamical structures that exist in the nonlinear regime. These halos exhibit a density distribution that can be characterized by a function of radius, known as the halo density profile. In other words, a halo density profile describes the distribution of dark matter within a galactic halo and can be used to predict the dark matter distribution within the host halo. 

Over the years, researchers have used various analytical methods and numerical simulations to develop halo density profiles that accurately represent observational data, particularly related to the rotation curves of galaxies \citep[see, e.g.,][]{1965TrAlm...5...87E, 1983MNRAS.202..995J, 1985MNRAS.216..273D, 1990ApJ...356..359H, 1993MNRAS.265..250D, 1996ApJ...462..563N}. The Navarro-Frenk-White (NFW) profile is one of the most commonly used density profiles for dark matter halos \citep{1996ApJ...462..563N}. The NFW profile is a two-parameter functional form that describes the halo mass distribution as a function of distance from the halo center. Another density profile that is obtained from analytical models is the Einasto profile, which is a generalization of a power law distribution with functionality of distance and shape parameter \citep{1965TrAlm...5...87E}. For both of the mentioned profiles, it is required that the logarithmic slope of the density distribution is $-2$ at the scale radius $r_{\rm s}$.

\subsection{Halo Concentration Parameter}
The concentration parameter is a dimensionless quantity that fundamentally represents the central density of galactic halos, determined by the ratio of the halo virial radius ($r_{\rm vir}$) to its scale radius ($r_{\rm s}$). The halo virial radius encompasses a volume where the average density is $200$ to $500$ times the critical density of the Universe. The concentration parameter holds significance in cosmology as it describes the shape of the halo density profile and the abundance of subhalos. Numerical simulations and analytical studies suggest that for precise predictions, the concentration parameter must dynamically change with halo mass and redshift \citep[e.g.,][]{2012MNRAS.423.3018P, 2014MNRAS.441.3359D, 2016MNRAS.460.1214L, 2016MNRAS.456.3068O}. This corresponds to the dynamics related to the merger history of dark matter halos and their evolutions. This correlation arises from the fact that smaller halos have already reached a state of virialization, leading to a higher concentration compared to host galactic halos. In this study, we use the concentration parameter defined in \cite{2016MNRAS.456.3068O} for calculations within the framework of GR. We also utilize the concentration parameters derived from \cite{2019MNRAS.487.1410M} and \cite{2021MNRAS.508.4140M} for computations related to the Hu-Sawicki $f(R)$ model of gravity and the nDGP model of gravity, respectively.

\subsection{Halo Mass Function}
The halo mass function plays a vital role in characterizing halos according to their mass. It offers an extensive insight into the distribution of dark matter halos with respect to mass. The density contrast is denoted as $\delta(x) \equiv [\rho(x)-\bar{\rho}]/\bar{\rho}$, where $\rho(x)$ signifies the density of a region denser than the average at point $x$, and $\bar{\rho}$ represents the mean density of the background. Essentially, the halo mass function serves as a crucial instrument for categorizing cosmological structures based on their mass, particularly identifying structures that surpass a specific density contrast threshold, which triggers their collapse and subsequent halo formation.

The precise forecasting of the halo mass function serves as a crucial evaluation of cosmological models. In this regard, the Press-Schechter (PS) method provides an analytical framework for modeling the statistics of hierarchical structure formation originating from initial density fluctuations \citep{1974ApJ...187..425P}. This method allows the formation of dark matter halos to be represented through stochastic processes and Gaussian density fields that are smoothed by a filter function. The PS formalism falls short in predicting halo abundances. For this purpose, several approaches have been carried out to overcome this issue \footnote{Besides, one of the most featured advantages is presented in the Sheth-Tormen (ST) formalism, which is grounded on triaxial collapse \citep{2001MNRAS.323....1S}. The ST model aligns more closely with simulations, yet it still overestimates high-mass halos. Although refinements have mitigated this overestimation, a certain level of discrepancy, particularly at high redshift, persists when compared to Bolshoi simulations \citep{2011ApJ...740..102K}.}.

The PS model was initially developed in \cite{1998A&A...337...96D}, where it was found that the collapse threshold is not a constant value but rather, it is dependent on the mass. Consequently, the collapse threshold can be characterized as
\begin{equation} \label{eqn:barrier}
\delta_{\rm cm} = \delta _{\rm c}(z) \left(1+\frac{\beta}{\nu^\alpha}\right),
\end{equation}
where $\alpha=0.585$, and $\beta=0.46$. In addition, $\nu=(\delta_{\rm c}/\sigma)^{2}$, where $\sigma(M, z)$ is the linear root-mean-square fluctuation of overdensities on a comoving scale with mass $M$ at redshift $z$. Using the excursion set formalism, one can demonstrate that Eq.\,(\ref{eqn:barrier}) corresponds to the following barrier
\begin{equation}\label{eqn:barrierv}
B(M)=\sqrt{a} \delta _{\rm c}(z) \left(1+\frac{\beta}{a \nu^\alpha} \right).
\end{equation} 
It is important to mention that this barrier, with a proper choice of $a$, yield a mass function that aligns with simulations \citep{1998A&A...337...96D}. The extended PS formalism can be enhanced by adjusting the barrier to include supplementary physical phenomena such as fragmentation, mergers, tidal torques, dynamical friction, and the cosmological constant. As demonstrated in \cite{2017JCAP...03..032D}, a modified threshold that is proportional to the initial barrier can accommodate these effects beyond spherical collapse, striving for a more precise model of halo formation:
\begin{eqnarray}\label{eqn:barrierf}
\delta_{\rm cm2} \simeq \delta_{\rm co} \left[1+\frac{\beta}{\nu^{\alpha}}+\frac{\Omega_{\Lambda}\beta_2}{\nu^{\alpha_2}}+\frac{\beta_3} {\nu^{\alpha_3}}\right]\;,
\end{eqnarray}
where $\alpha=0.585$, $\beta=0.46$, $\alpha_2=0.4$, $\beta_2=0.02$, $\alpha_3=0.45$, and $\beta_3=0.29$. Also, $\Omega_{\Lambda} \simeq 0.692$ is the density parameter of cosmological constant.

In the framework of the excursion set theory, the mass function is defined as the comoving number density of halos within a particular mass interval \citep{1991ApJ...379..440B}:
\begin{equation}\label{eqn:universal}
n(M,z)=\frac{\overline{\rho}}{M^{2}}\left|\frac{d\log{\nu }}{d\log M}\right|\nu f(\nu)\;,
\end{equation}
where $f(\nu)$ is known as multiplicity function that denotes the distribution of the first crossing.

The subsequent relation can be derived by considering the barrier provided by Eq.\,(\ref{eqn:barrierf}) and incorporating the influences of angular momentum, dynamical friction, and cosmological constant, as demonstrated in \cite{2017JCAP...03..032D}:
\begin{equation}\label{nufnu}
\nu f(\nu) \simeq A_{2}\sqrt{\frac{a\nu}{2\pi}}l(\nu)\exp{\left\{-0.4019 a \nu^{2.12} m(\nu)^{2}\right\}},
\end{equation}
where
\begin{eqnarray}
& l(\nu)=\left(1+\frac{0.1218}{\left(a\nu\right)^{0.585}}+\frac{0.0079}{\left(a\nu\right)^{0.4}}+\frac{0.1}{\left(a\nu\right)^{0.45}}\right),\nonumber\\~\\
& m(\nu)=\left(1+\frac{0.5526}{\left( a\nu \right)^{0.585}} +\frac{0.02}{\left(a\nu\right)^{0.4}}+\frac{0.07}{\left(a\nu\right)^{0.45}}\right)\nonumber.
\end{eqnarray}
Also, the constants are assigned specific values, namely $A_2=0.93702$ and $a=0.707$. By substituting Eq.\,(\ref{nufnu}) into Eq.\,(\ref{eqn:universal}), a precise semianalytical mass function can be derived, referred to as the mass function of GR henceforth.

Nonetheless, in MG theories, it is necessary to rescale the mass function. This is because, in those models that introduce an additional scalar degree of freedom, Birkhoff’s theorem is violated. This results in the collapse process being contingent on the environment, a phenomenon known as the chameleon screening mechanism. In this mechanism, dense regions partially mitigate the amplified gravitational forces. Consequently, the threshold of surplus locations depends on the halo mass, redshift, and the particular configuration of the model parameters, which alters the gradients of the scalar field.

In \cite{2022PhRvD.105d3538G}, a rescaled halo mass function for the Hu-Sawicki $f(R)$ gravity is presented based on $N$-body simulations:
\begin{equation}\label{massfuncfr}
\frac{n_{f(R)}}{n_{\rm GR}}= 1+ A\exp\left[\frac{\left(X-B\right)^{2}}{C^{2}}\right],
\end{equation}
where $X\equiv\ln(\sigma^{-1})$, which naturally depends on $M_{\rm vir}$ and $z$. Also, based on varoius models of Hu-Sawichi $f(R)$ gravity, the best values of $A, B$, and $C$ are provided in Table \ref{tab1}. 

Furthermore, in \cite{2021MNRAS.508.4140M}, a suitable form for the halo mass function in nDGP models is proposed, employing dark matter only simulations:
\begin{eqnarray}\label{massfuncndgp}
\frac{n_{\rm nDGP}}{n_{\rm GR}}= 1+ D(H_{0}r_{\rm c})\times\hspace*{3.5cm}\nonumber\\
\left[\tanh\left(\log(M_{\rm vir}/M_{\odot}h^{-1})-E(z)\right)+F(z)\right],
\end{eqnarray}
where
\begin{eqnarray}
& D(H_{0}r_{\rm c}) = (0.342 \pm 0.014)(H_{0}r_{\rm c})^{-1}, \nonumber\\\nonumber ~\\
& E(z) = (14.87 \pm 0.03) - (0.481 \pm 0.01)z, \\\nonumber ~\\
& F(z) = (0.864 \pm 0.008) + (0.047 \pm 0.005)z. \nonumber
\end{eqnarray}
It is important to highlight that the aforementioned parameters represent the best-fit values, derived via the method of unweighted least squares.
\begin{table}
\caption{Best-fit values of parameters in Eq.\,(\ref{massfuncfr}) for the Hu-Sawicki $f(R)$ models \citep{2022PhRvD.105d3538G}.}
\begin{tabular}{|c|c|c|c|} 
\hline 
\hline
Model & A & B & C \\ 
\hline
\hline
f4 & 0.630 & 1.062 & 0.762 \\ 
\hline
f5 & 0.230 & 0.100 & 0.360 \\ 
\hline
f6 & 0.152 & -0.583 & 0.375 \\ 
\hline
\hline
\end{tabular}
\label{tab1}
\end{table}
%%%%%%%%%%%%%%%%%%%%%%%%%%%%%%%%%%%%%%%%%%%%%%%%%%%%%%%%%%%%%%%%
\section{PBH-NS Merger Rate}\label{sec:iv}
Consider a scenario where a PBH with mass $m_1$ and a NS with mass $m_2$ are situated within an isolated dark matter halo. Assume they abruptly cross paths on a hyperbolic trajectory, and their relative velocity at a large distance is denoted by $v_{\rm rel}=|v_1-v_2|$. Given this premise, two-body scattering suggests that significant gravitational radiation is emitted at the nearest physical separation, referred to as the periastron. If the radiated gravitational energy surpasses the system’s kinetic energy, the compact objects will become gravitationally bound, resulting in a binary formation. The upper limit of the periastron is dictated by this condition \citep{123456789p}:
\begin{equation}\label{periast0}
r_{\rm mp} = \left[\frac{85 \pi}{6\sqrt{2}}\frac{G^{7/2}m_{1}m_{2}(m_{1}+m_{2})^{3/2}}{c^{5}v_{\rm rel}^{2}}\right]^{2/7},
\end{equation}
where $G$ represents the gravitational constant and $c$ denotes the speed of light. Furthermore, within the framework of the Newtonian approximation, the impact parameter can be defined as \citep{2018CQGra..35f3001S}:
\begin{equation}\label{impact}
b^{2}(r_{\rm p}) = \frac{2G(m_{1}+m_{2})r_{\rm p}}{v_{\rm rel}^{2}} + r_{\rm p}^{2}.
\end{equation}

\begin{figure*}
\centering
\includegraphics[width=0.45\textwidth]{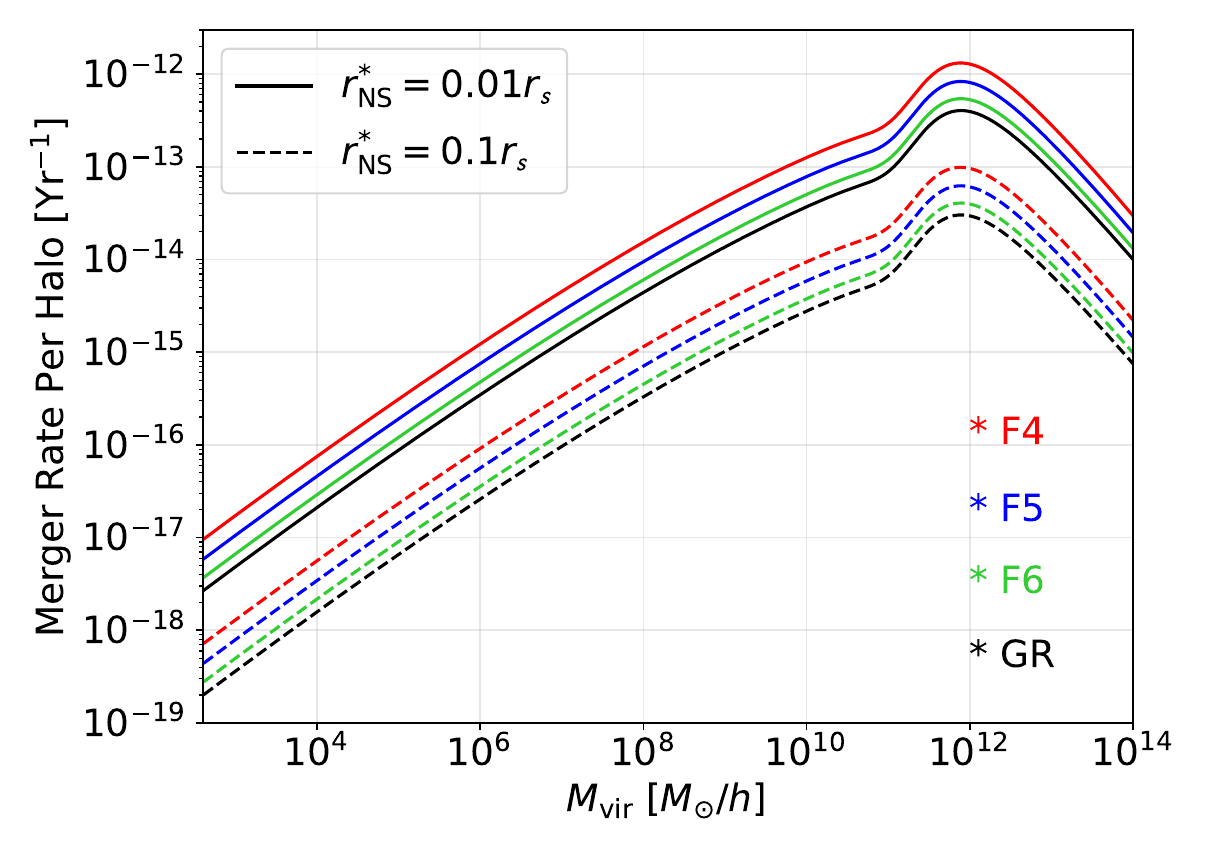}
\includegraphics[width=0.45\textwidth]{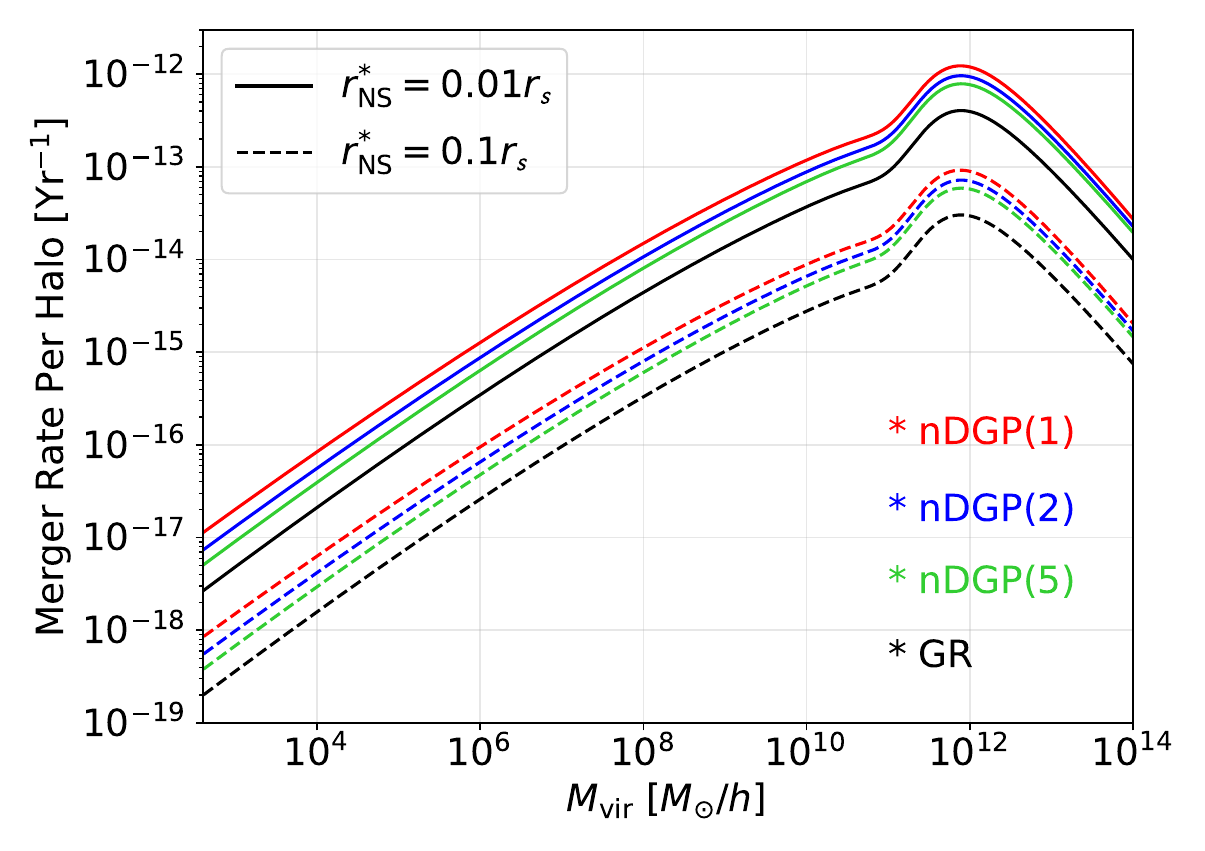}
\caption{The merger rate of PBH-NS binaries within each halo as a function of halo mass in the present-time Universe, considering Hu-Sawicki $f(R)$ gravity (left) and nDGP gravity (right), for various parameter values of the model. GR predictions are over-plotted for comparison. Solid and dashed lines correspond to $r^{*}_{\rm NS}=0.01 r_{\rm s}$, and $0.1 r_{\rm s}$, respectively. The halo density is assumed to follow the NFW profile. We have assumed $M_{\rm PBH}=5\,M_{\odot}$, $M_{\rm NS}=1.4\,M_{\odot}$, and $f_{\rm PBH}=1$.}
\label{fig_1}
\end{figure*}

Furthermore, when the strong limits of gravitational focusing are set, that is, $r_{\rm p}\ll b$, the tidal forces exerted by the nearby compact objects on the binary can be disregarded. Consequently, the cross-section for the binary formation is given by \citep{1989ApJ...343..725Q, 2002ApJ...566L..17M}
\begin{equation}\label{crosssec}
\xi(m_1,m_2,v_{\rm rel})=\pi b^2(r_{\rm mp})\simeq\frac{2\pi G(m_1+m_2)r_{\rm mp}}{v^2_{\rm rel}}.
\end{equation}
Therefore, the rate at which binaries form within each galactic halo conforms to the following equation \citep{2016PhRvL.116t1301B}:
\begin{equation}\label{per_mer}
\Gamma =4\pi \int_{0}^{r_{\rm vir}} \left(\frac{f_{\rm PBH}\rho_{\rm h}}{m_1}\right) \left(\frac{\rho_{\rm NS}}{m_2}\right) \langle\xi v_{\rm rel}\rangle \, r^2\,dr,
\end{equation}
where $0 < f_{\rm PBH} \leq 1$ represents the fraction of PBHs determining their contribution to dark matter, $\rho_{\rm h}$ denotes the halo density profile, which can be modeled using the NFW or Einasto profiles. The angle bracket is an average over the PBH relative velocity distribution within the galactic halo. Additionally, $\rho_{\rm NS}$ represents the density profile of NSs, characterized by the following spherically-symmetric form:
\begin{equation}\label{nsdensity}
\rho_{\rm NS}(r)= \rho^{*}_{\rm NS}\exp\left(-\frac{r}{r^{*}_{\rm NS}}\right)
\end{equation}
The parameters $\rho^{*}_{\rm NS}$ and $r^{*}_{\rm NS}$ represent the characteristic density and radius of NSs, respectively. Eq.\,(\ref{nsdensity}) illustrates that to determine the distribution of NSs, two quantities must be specified. Firstly, the defining radius of a NS, $r^{*}_{\rm NS}\simeq (0.01\mbox{-}0.1)r_{\rm s}$, is considered \citep{2022ApJ...931....2S}. Secondly, the defining density of NSs, $\rho^{*}_{\rm NS}$, is established by normalizing the NS distribution to their projected population within a specific galaxy. In order to achieve this, we employ the time-constant form of the initial Salpeter stellar mass-function, denoted as $\chi(m_{*})\sim m_{*}^{-2.35}$. Moreover, we posit that all stars within the mass range of $M_{\rm NS}\simeq (8\mbox{-}20)\,M_{\odot}$ undergo supernova explosions, leading to the formation of NSs. As a result, the number count of NSs within a galaxy that has a stellar mass $M_{*}$ is determined as follows:
\begin{equation}
n_{\rm NS}=M_{*}\int_{m_{*}^{\rm min}}^{{m_{*}^{\rm max}}}\chi(m_{*})dm_{*},
\end{equation}
where $\chi(m_{*})m_{*}$ is normalized to unity. It is crucial to determine the galactic stellar mass by defining the correlation between stellar mass and halo mass, represented as $M_{*}(M_{\rm vir})$. To achieve this, we employ the stellar mass-halo mass relationship outlined in \cite{2013ApJ...770...57B}, under the assumption that the highest concentration of NSs is located in the central region of the galactic halo.

The calculation of the merger rate of PBH-NS binaries also depends on dark matter particle velocity dispersion within galactic halos. Recent studies suggest that MG model velocity dispersion profiles bear a close resemblance to those in the conventional cosmological model \citep{2015PhRvL.115g1306H}. Consequently, it is justifiable to apply the dark matter particle velocity dispersion relation, as derived in \cite{2012MNRAS.423.3018P} to our analysis. Furthermore, we set the probability distribution function of relative velocities among PBHs in the galactic halos to be regulated by the Maxwell–Boltzmann distribution, which is truncated at the virial velocity \citep{2021PhRvD.103l3014F, 2022PhRvD.105d3525F}.

\begin{figure*}
\centering
\includegraphics[width=0.45\textwidth]{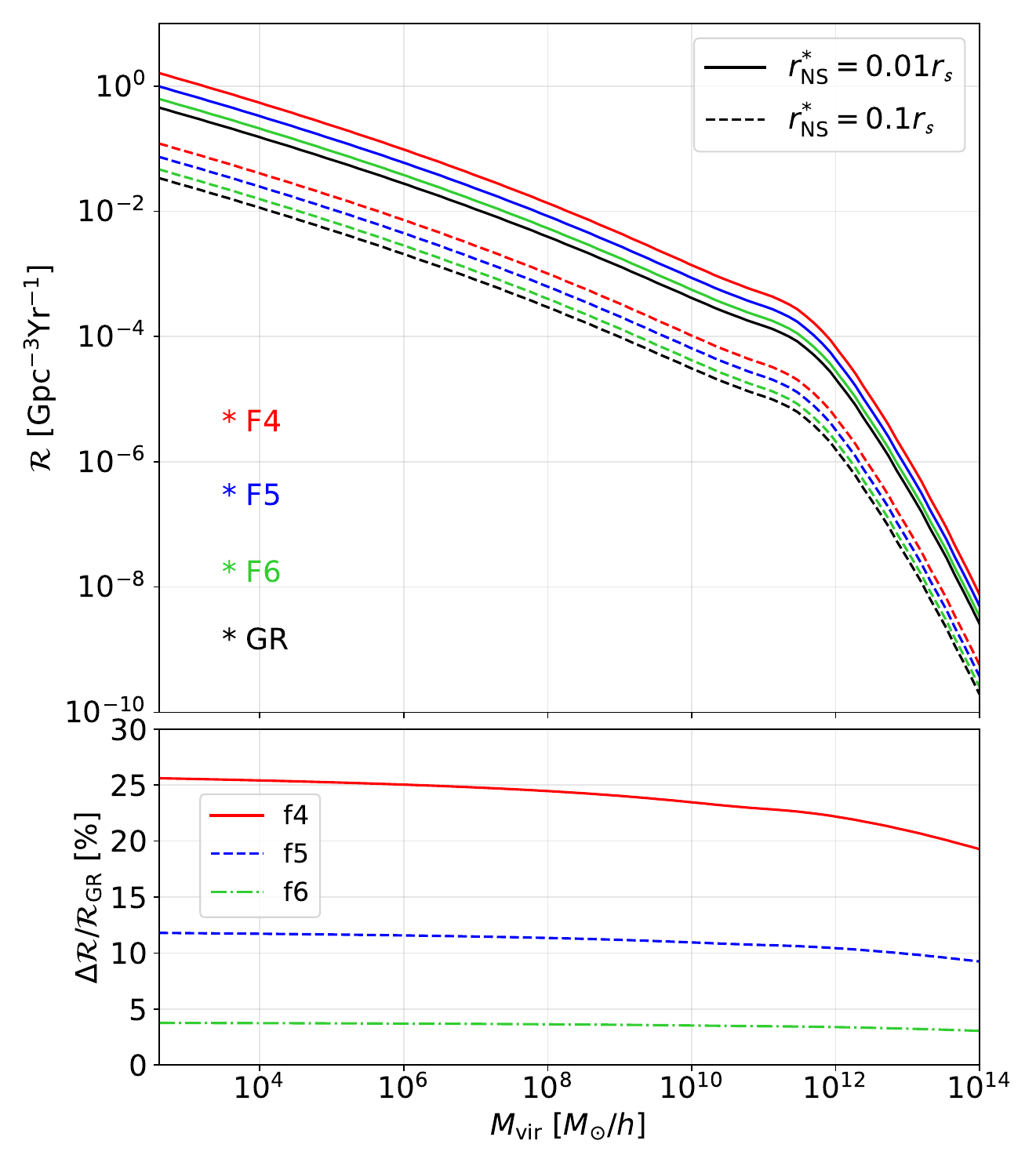}
\includegraphics[width=0.45\textwidth]{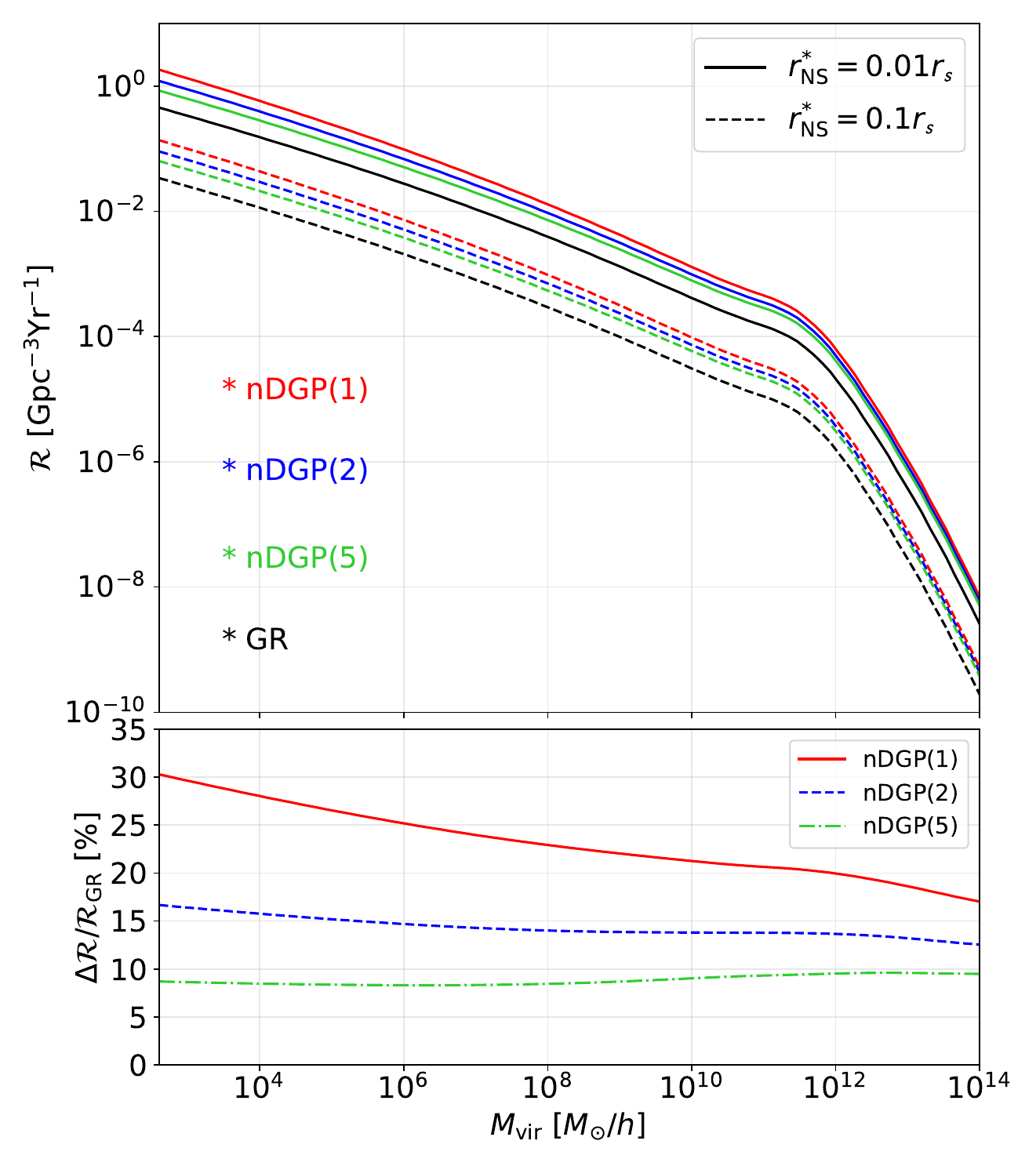}
\caption{Top: The total merger rate of PBH-NS binaries as a function of halo mass in the present-time Universe, for Hu-Sawicki $f(R)$ (left) and nDGP gravities (right), and with the NFW density profile for the halos, compared to GR predictions. Solid and dashed lines correspond to $r^{*}_{\rm NS}=0.01 r_{\rm s}$, and $0.1 r_{\rm s}$, respectively. 
Bottom: The relative enhancement of the merger rate from MG predictions to those from GR as a function of halo mass. We have assumed $M_{\rm PBH}=5\,M_{\odot}$,  $M_{\rm NS}=1.4\,M_{\odot}$, and $f_{\rm PBH}=1$.}
\label{fig_2}
\end{figure*}

To calculate the total merger rate of PBH-NS binaries, one can multiply the halo mass function, ${\rm d}n(M)/dM$, by the rate of binary formation within each halo, $\Gamma(M)$, and integrate over a minimum mass of halos:
\begin{equation}\label{tot_mer}
\mathcal{R}=\int_{M_{\rm min}}\frac{dn}{dM_{\rm vir}}\Gamma(M_{\rm vir})dM_{\rm vir}.
\end{equation}
The above formula indicates that the ultimate result is minimally influenced by the upper limit. This can be attributed to the fact that the halo mass function incorporates a term that decreases exponentially, rendering the merger rate to decrease as the halo mass increases. Conversely, the lower limit holds significant importance. This concurs with the dynamics of hierarchical structure formation, given that the smallest halos are anticipated to exhibit greater density than the host halos. Nonetheless, dynamical relaxation processes lead to the evaporation of the smallest halos by ejecting objects. Furthermore, merger and accretion processes lose their effectiveness during the dark energy dominated era. Consequently, the smallest halos could have entirely evaporated in the past $3$ Gyr, leading to the potential disregard of signals from some halo mergers \citep{2016PhRvL.116t1301B}. Hence, the evaporation timeframe of dark matter halos guarantees their presence in the present-time Universe. In our prior studies, we have established that the evaporation period of dark matter halos with a standard mass of $400\,M_{\odot}$, hosting PBHs with a mass of $30\,M_{\odot}$, is approximately $3$ Gyr \citep{2021PhRvD.103l3014F, 2022PhRvD.105d3525F}. As a result, one can consider dark matter halos with the aforementioned typical mass as the minimum mass in our analysis. However, dark matter halos hosting PBHs lighter than $30\,M_{\odot}$ can possess slightly lower masses than $400\,M_{\odot}$ to persist today, while those hosting PBHs heavier than $30\,M_{\odot}$ can have slightly higher masses than $400\,M_{\odot}$ to fulfill the conditions.
%%%%%%%%%%%%%%%%%%%%%%%%%%%%%%%%%%%%%%%%%%%%%%%%%%%%%%%%%%%%%%%%
%%%%%%%%%%%%%%%%%%%%%%%%%%%%%%%%%%%%%%%%%%%%%%%%%%%%%%%%%%%%%%%%
\section{Results and Discussions}\label{sec:v}
%%%%%%%%%%%%%%%%%%%%%%%%%%%%%%%%%%%%%%%%%%%%%%%%%%%%%%%%%%%%%%%%
Fig.\,\ref{fig_1} illustrates the merger rate of PBH-NS binaries within each halo as a function of halo mass for Hu-Sawicki $f(R)$ and nDGP gravities with different parameter values, compared with GR predictions. We have employed the NFW density profile and assumed that PBHs have a mass of $5\,M_{\odot}$ and NSs have a mass of $1.4\,M_{\odot}$ for this calculation. Moreover, we have considered that PBHs account for all of the dark matter, that is, $f_{\rm PBH} = 1$. In this figure, calculations have been performed for the permissible boundaries of the characteristic radius of neutron stars, i.e., $r^{*}_{\rm NS}=0.01$, and $0.1\,r_{\rm s}$. As expected, the respective outcomes for any other potential values lie between these two limits. The left subfigure indicates that the merger rate of PBH-NS binaries in each halo is greater for all Hu-Sawicki $f(R)$ models than for GR. This is attributed to the nonlinear effects of $f(R)$ gravity on density fluctuations and halo formation and evolution, which are incorporated in all stages of the calculations. Another notable finding is that the impact of the field strength on density fluctuations diminishes progressively from f4 to f6. This is because the merger rate of PBH-NS binaries is dependent not only on density fluctuations but also on the field strength $f_{R0}$. A similar result can be found by looking at the right subfigure. Accordingly, the merger rate of PBH-NS binaries in each halo in nDGP models is much higher than the results obtained for GR. It is also clear that the enhancement of the merger rate is highest for the nDGP(1) model, moderate for the nDGP(2) model and lowest for the nDGP(5) model.

\begin{figure*}
\centering
\includegraphics[width=0.45\textwidth]{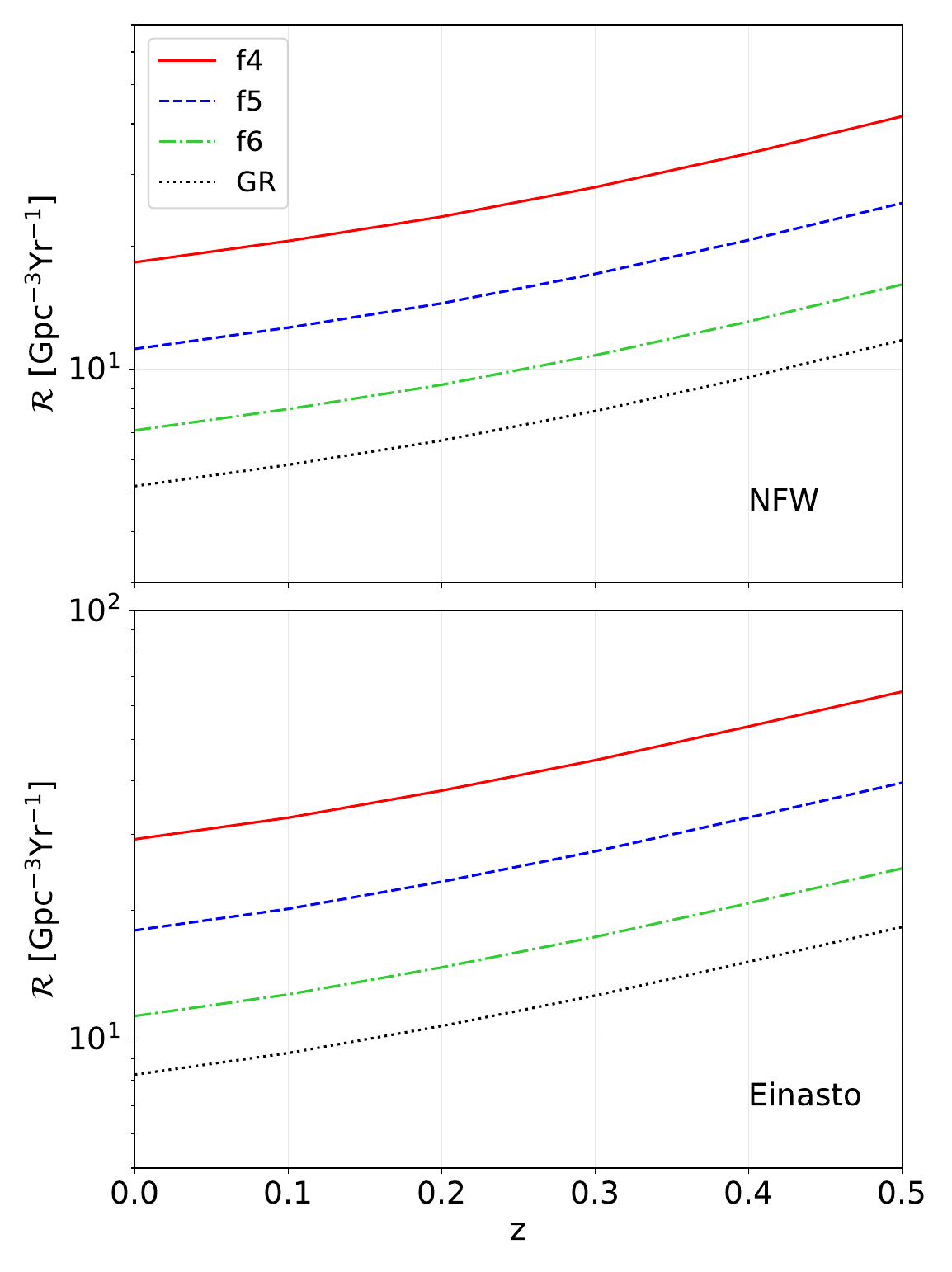}
\includegraphics[width=0.45\textwidth]{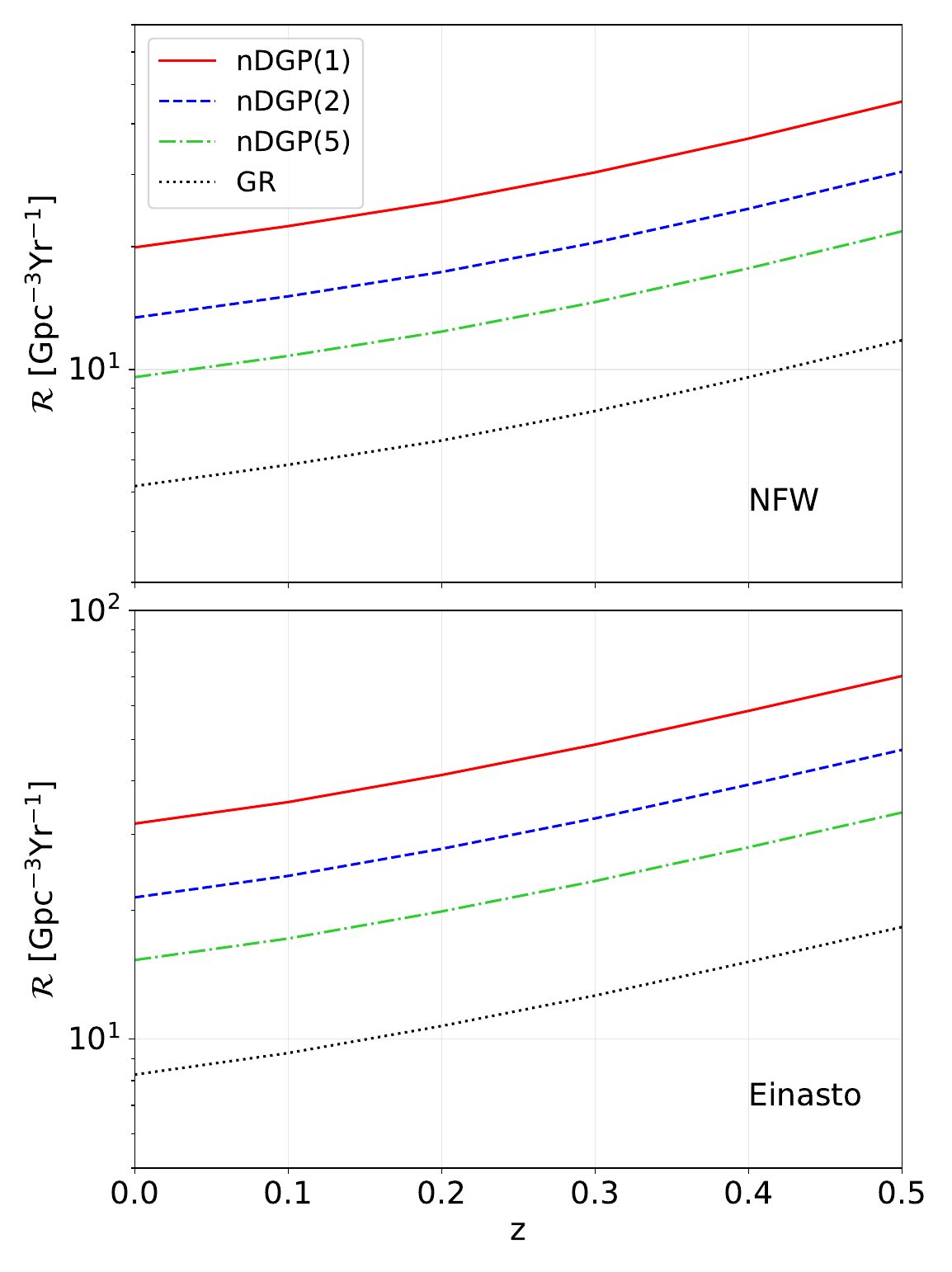}
\caption{Upper limit of the merger event rate of PBH-NS binaries as a function of redshift, considering Hu-Sawicki $f(R)$ gravity (left) and nDGP gravity (right). These results are compared with those obtained in GR. Both NFW and Einasto density profiles are employed. We have assumed $M_{\rm PBH}=5\,M_{\odot}$,  $M_{\rm NS}=1.4\,M_{\odot}$, and $f_{\rm PBH}=1$.}
\label{fig_3}
\end{figure*}

In Fig.\,\ref{fig_2}, we have shown the merger rates of PBH-NS binaries per unit time and volume in Hu-Sawicki $f(R)$ and nDGP models, and compared them with the GR results using NFW density profile. These calculations are conducted for the present-time Universe. We have applied Eqs.\,(\ref{nufnu}), \,(\ref{massfuncfr}), and \,(\ref{massfuncndgp}) for the halo mass functions of GR, Hu-Sawicki $f(R)$, and nDGP models, respectively. We have also utilized the relations derived in \citep{2016MNRAS.460.1214L, 2022PhRvD.105d3538G, 2021MNRAS.508.4140M} as the concentration parameters correspondingly. The merger rate of PBH-NS binaries decreases as the minimum mass of dark matter halos increases. This means that the binary merger rate is more affected by smaller halos than larger ones. These results support the dynamics of hierarchical structures, which imply that the mass of halos and the density of dark matter particles are inversely related. Therefore, the subhalos have more dark matter particles and higher concentration than the host halo. It is noteworthy that the merger rate of PBH-NS binaries can be precisely determined by integrating across the area under the curves. Our analysis reveals that for any Hu-Sawicki $f(R)$ model, the merger rate of PBH-NS binaries is higher than that of GR. Accordingly, the merger rate of PBH-NS binaries clearly depends on the field strength $f_{R0}$. As the field strength becomes smaller, the merger rate approaches the GR result. Based on this argument, one can utilize the value of the field strength and GW data to compare theoretical predictions with observations and estimate the number of PBHs. This approach holds potential as a new technique.

A similar pattern for the nDGP models can also be obtained. The difference between the merger rate of PBH-NS binaries in nDGP gravity and GR is more pronounced for nDGP(1) than for nDGP(2) and nDGP(5) respectively. This means that the nDGP models have different effects on the merger rate of PBH-NS binaries depending on the parameter $n=H_{0}r_{\rm c}$, which controls the strength of the extra dimension. We have also quantified the amplification in the merger rate of PBH-NS binaries for both gravitational models. The two models yield different reinforcement processes. Nevertheless, the growth of the merger rate for both models is inversely related to the minimum mass of halos, leading to the most substantial growth occurring in low-mass halos. Furthermore, it is apparent that the enhancement in the merger rate of PBH-NS binaries is slightly greater in the nDGP models compared to the results obtained from Hu-Sawicki $f(R)$ models.

\begin{figure*}
\centering
\includegraphics[width=0.45\textwidth]{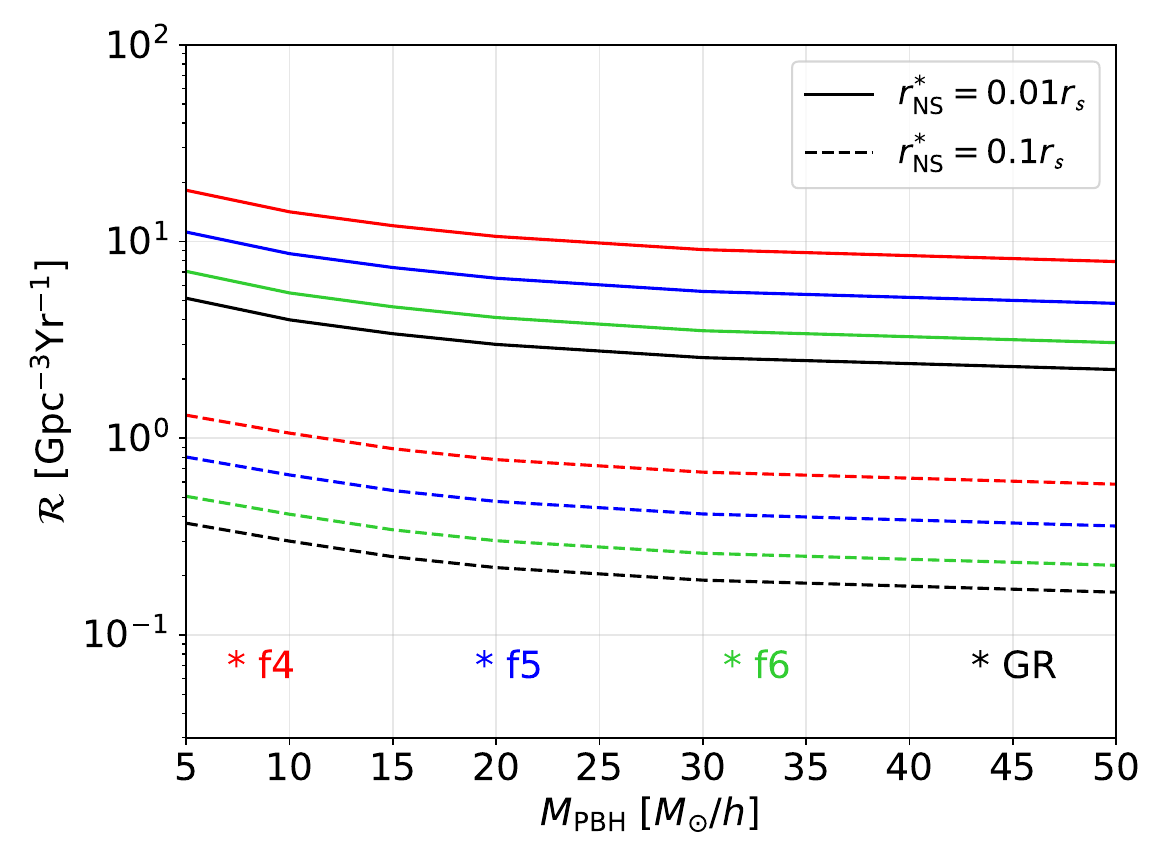}
\includegraphics[width=0.45\textwidth]{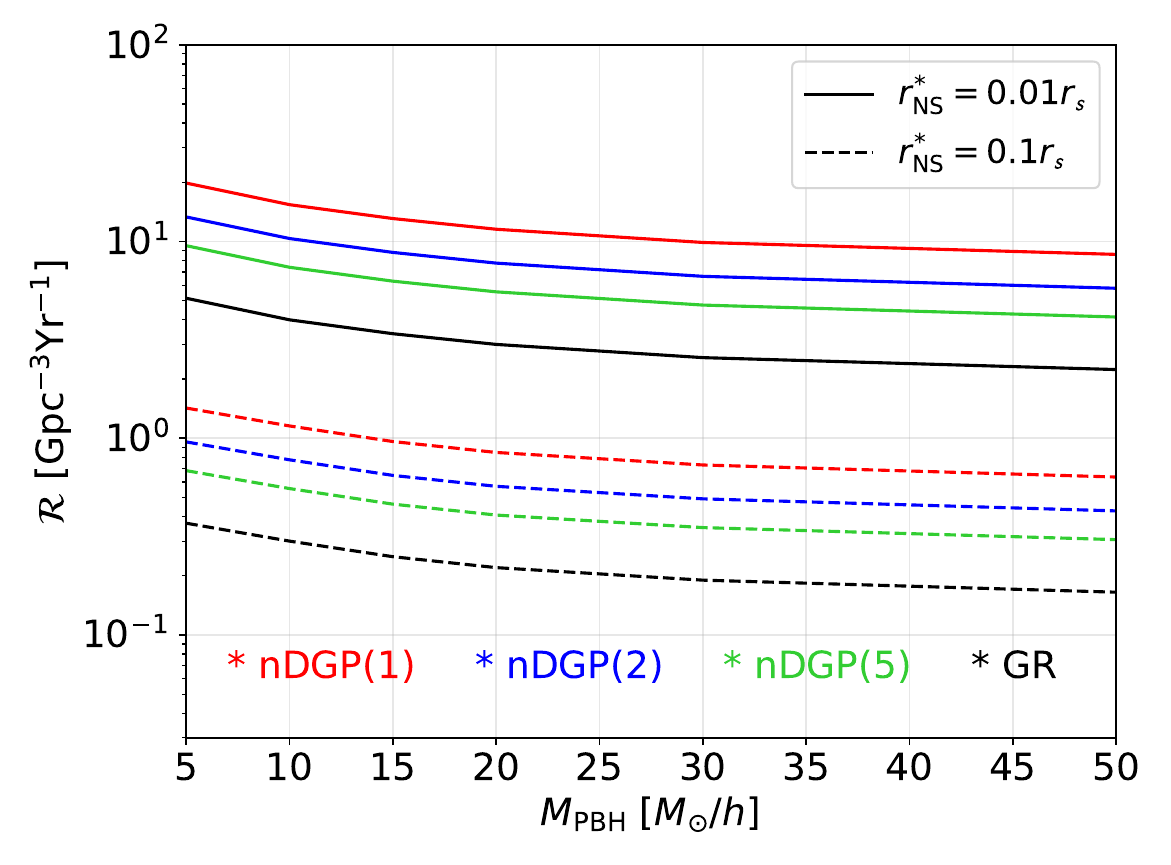}
\caption{Total merger rate of PBH-NS binaries as a function of different PBH masses, considering Hu-Sawicki $f(R)$ gravity (left) and nDGP gravity (right). These results are compared with those obtained in GR. Solid and dashed lines correspond to $r^{*}_{\rm NS}=0.01 r_{\rm s}$, and $0.1 r_{\rm s}$, respectively. The NFW density profile is employed. We have assumed $M_{\rm NS}=1.4\,M_{\odot}$, and $f_{\rm PBH}=1$.}
\label{fig_4}
\end{figure*}

\begin{figure*}
\centering
\includegraphics[width=0.45\textwidth]{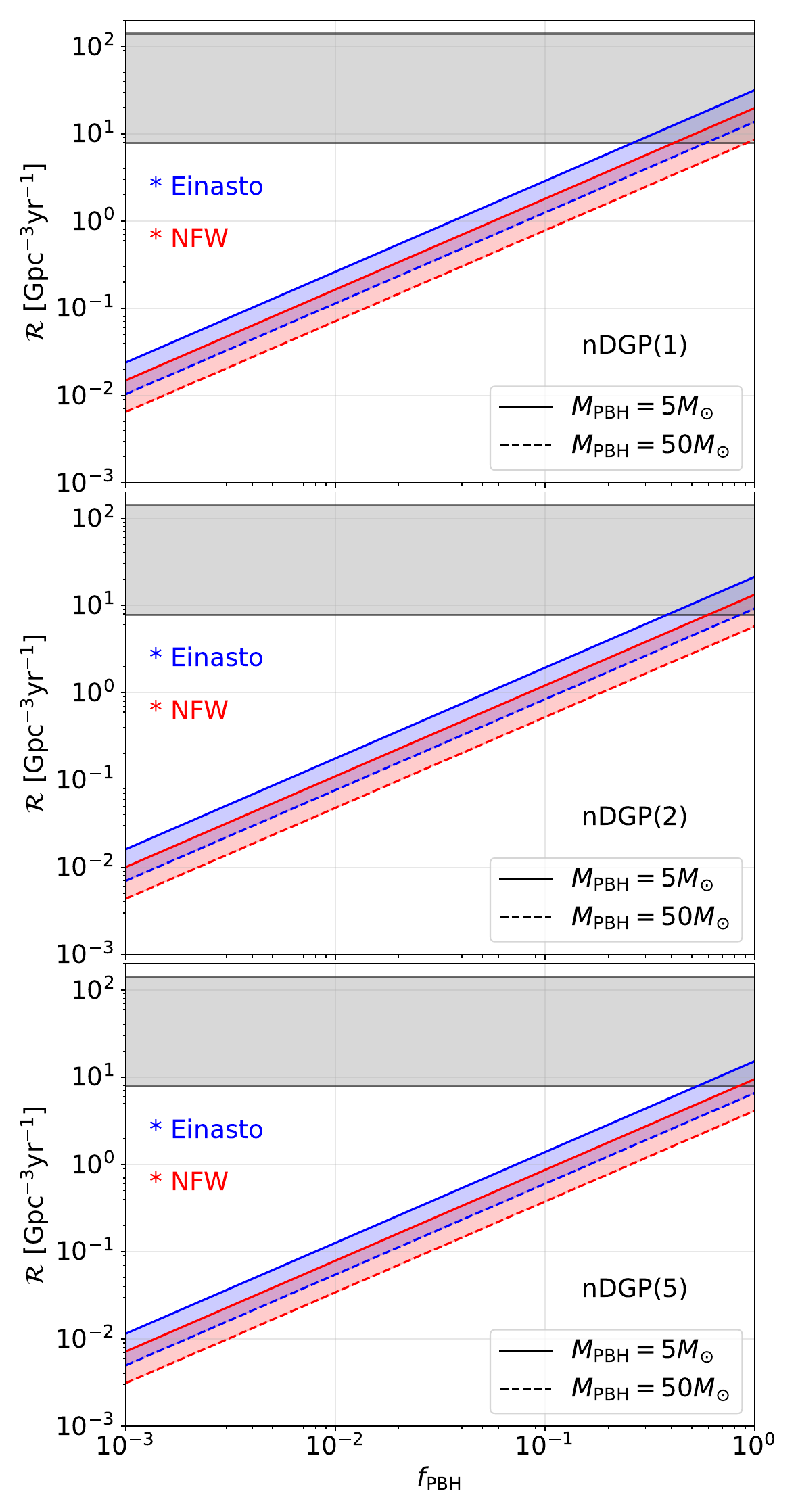}
\includegraphics[width=0.45\textwidth]{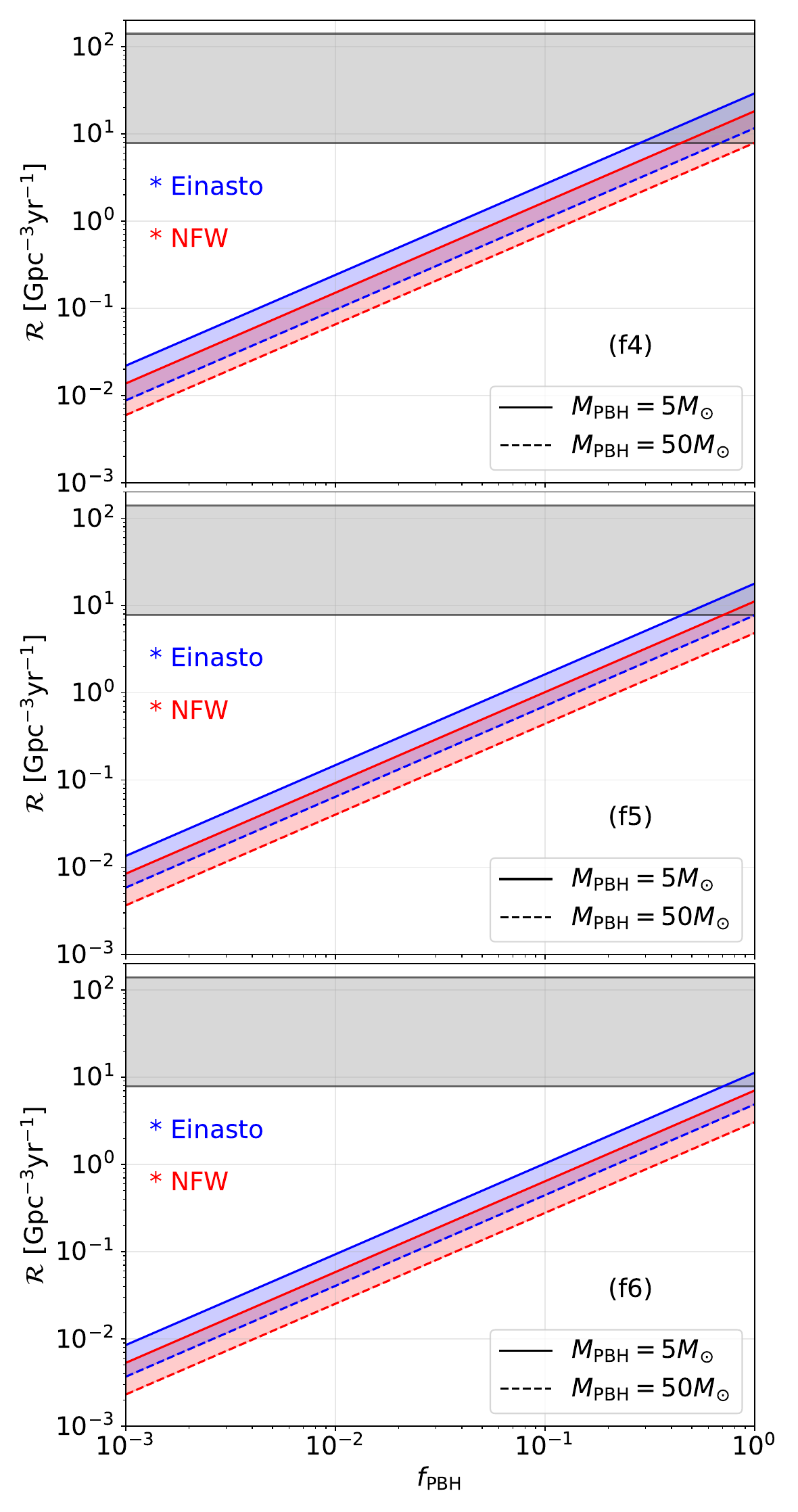}
\caption{Upper limit on the total merger event rate of PBH-NS binaries with respect to the fraction and mass of PBHs, considering Hu-Sawicki $f(R)$ gravity (left) and nDGP gravity (right). The shaded gray bands represent the total merger rate of BH-NS binaries, as estimated by the LIGO-Virgo-KAGRA detectors during their latest observing run, within the range of $(7.8-140)\,{\rm Gpc^{-3} yr^{-1}}$. The shaded red and blue bands show the results for the NFW and Einasto density profiles, respectively. We have assumed $M_{\rm NS}=1.4\,M_{\odot}$.}
\label{fig_5}
\end{figure*}

The specific goal of this analysis is to illustrate the merger rate of PBH-NS binaries throughout the evolution of the Universe, facilitating comparison with the merger events detected by the LVK observatories. This effort is grounded in gravitational models and aims to offer theoretical forecasts, elucidating the prospective trajectory of GW phenomena and the progression of these detection instruments.. The current sensitivity of GW detectors allows the detection of merger events within a comoving volume of $50\, {\rm Gpc^{3}}$, corresponding to a redshift of roughly $0.75\,z$. This presents a challenging issue in calculating the redshift evolution of the merger rate of PBH-NS binaries. The sensitivity to changes in redshift is emphasized, particularly in Eq.\,(\ref{tot_mer}), which includes the halo mass function and the concentration parameter. In Fig\,\ref{fig_3}, we have depicted the redshift evolution of the merger rate of PBH-NS binaries for Hu-Sawicki $f(R)$ and nDGP models, contrasting the results with those for GR while taking into account NFW and Einasto density profiles. The direct relation between the merger rate of PBH-NS binaries and redshift is evident, which can be attributed to the influence of hierarchical dynamics and halo merger structure. The findings suggest that the redshift evolution of the merger rate PBH-NS binaries in the mentioned gravitational models is higher than that obtained from the GR framework, potentially leading to an amplified binary merger rate while going back in time. Furthermore, one has to highlight the effect of model parameters, like $f_{R0}$ and $n=H_{0}r_{\rm c}$, on the increase of the merger rate of PBH-NS binaries during the late-time Universe. Furthermore, the merger rates derived from all models, considering the Einasto density profile, display an increase of approximately $60\%$ compared to those obtained when considering the NFW density profile.

Fig.\,\ref{fig_4} presents the merger rate of PBH-NS binaries for both Hu-Sawicki $f(R)$ and nDGP models of gravity as a function of PBH mass and the two characteristic radii of NS, compared with GR. We have only provided the results for the NFW density profile, knowing the proportionality of the results between the two profiles. Also given the linear scaling of the merger rates with PBH abundance, we have only shown the results for $f_{\rm PBH}=1$. In all the models considered, the merger rate of PBH-NS binaries has an inverse relationship with the mass of PBHs. This is because the number density of PBHs changes inversely with their mass within a certain volume. Smaller PBHs are more likely to participate in PBH-NS binary formations than larger ones. This is because the formation of PBH-NS binaries is influenced by factors such as the mass range of PBHs and the density contrast boost factor \citep{2021JCAP...10..019T}.

In our previous discussions, we assumed PBHs can provide the maximum contribution to dark matter. Nevertheless, it is interesting to determine the merger rate of PBH-NS binaries, considering various assumptions about PBH fractions and masses. To this end, in fig.\,\ref{fig_5}, we have depicted the upper limit of the cumulative merger rate of PBH-NS binaries as a function of the mass spectrum of stellar mass PBHs for both Hu-Sawicki $f(R)$ and nDGP models. This calculation incorporates both NFW and Einasto density profiles. It is once again clear that the merger rate of PBH-NS binaries is inversely related to the mass of PBHs.

Moreover, the findings indicate that the Hu-Sawicki $f(R)$ and nDGP theoretical models enforce more stringent constraints on the fraction of PBHs, particularly when smaller masses of PBHs contribute to binary systems. The figure also suggests that the value of the field strength $f_{R0}$ plays a significant role in constraining the fraction of PBHs. This reveals a direct relation between the field strength and the severity of the constraints on a fraction of PBHs. In this context, the most rigorous constraints can be derived from the $f(R)$
gravity with the field strength f4, f5, and f6 respectively. Similar results are inferred from nDGP models, wherein the most stringent constraints are imposed on the abundance of PBHs from nDGP(1), nDGP(2), and nDGP(5) models, respectively.

Additionally, it can be deduced that the most precise constraint in this study is derived from the $f(R)$ models, specifically when the Einasto density profile and $M_{\rm PBH}\simeq 5\,M_{\odot}$ are considered. A similar outcome is obtained in the nDGP model with the same PBH mass and density profile, particularly when the nDGP(1) model is taken into account. Despite the presence of theoretical uncertainties, if the proportion of PBHs comprises roughly over $10\%$ of the total dark matter content, the findings align with the LVK sensitivity band.

\begin{figure}
\includegraphics[width=\linewidth]{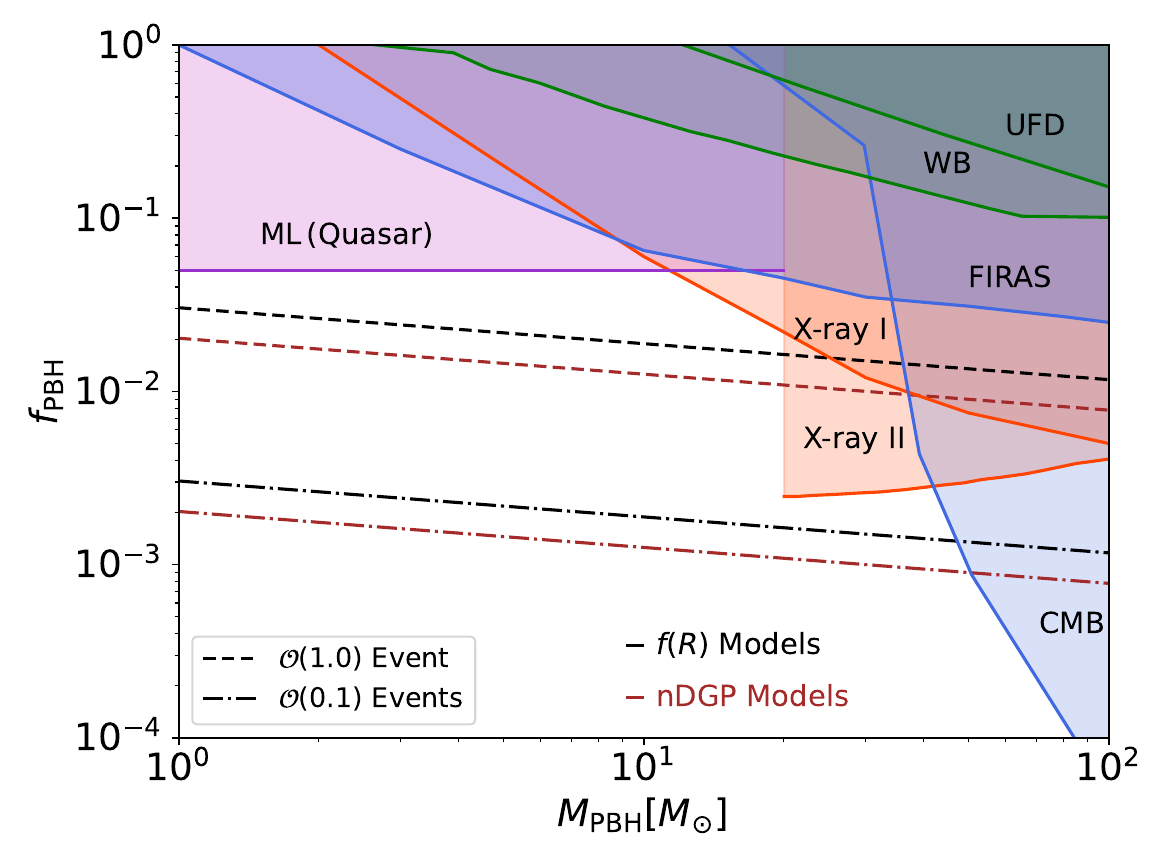}
\caption{The expected upper bounds on PBH fraction as a function their mass, for Hu-Sawicki $f(R)$ (black lines) and nDGP (brown lines) gravities. The dashed and dot-dashed lines represent scenarios for which $\mathcal{O}(1)$ and $\mathcal{O}(0.1)$ annual detections of PBH-NS merger events are expected in the comoving volume of $50\,{\rm Gpc}^{3}$, with $M_{\rm PBH}=5 M_{\odot}, M_{\rm NS}=1.4 M_{\odot}$. Non-GW constraints on PBH dark matter, including gravitational microlensing constraints from quasars (ML Quasar), constraints from the disruption of wide binaries (WB), constraints from the disruption of ultra-faint dwarfs (UFD), X-ray constraints related to accreting PBHs (X-ray I), and the corresponding one in the Milky Way (X-ray II), and constraints from modification of the cosmic microwave background (CMB) spectrum due to accreting PBHs, including particle dark matter accretion, as well as the FIRAS data are provided as benchmarks for this analysis.}
\label{fig_6}
\end{figure}

Nonetheless, it is important to highlight that the constraints derived from the merger rate of PBH-NS binaries in these models signify the upper limits allowed by GW detectors. This is attributed to the fact that the BH component in the mergers, which originate from astrophysical sources, can also be detected by the LVK detectors. Our analysis suggests that, to annually observe a minimum of $\mathcal{O}(10)$, $\mathcal{O}(1)$, and $\mathcal{O}(0.1)$ merger events of PBH-NS binaries with a mass of $(M_{\rm PBH}=5\,M_{\odot}, M_{\rm NS}=1.4\,M_{\odot})$ in the comoving volume of $50\,{\rm Gpc^3}$, the anticipated upper bounds on the abundance of PBHs in both gravitational models are estimated to be $f_{\rm PBH} \sim \mathcal{O}(10^{-1})$, $\mathcal{O}(10^{-2})$, and $\mathcal{O}(10^{-3})$, respectively.

However, there exist robust constraints from non-GW data on the abundance of PBHs within the mass range of $5\mbox{-}50\,M_{\odot}$. In Figure \ref{fig_6}, we have shown a comparative analysis between the outcomes of $f(R)$ and nDGP models concerning the upper bounds on the abundance of PBHs and the non-GW data related to the examined mass range. 
The non-GW data include constraints from gravitational microlensing of quasars (ML Quasar) and OGLE-II+III, disruption of wide binaries (WBs), disruption of ultrafaint dwarf galaxies (UFDs), X-ray constraints associated with accreting PBHs (X-ray I) and its Milky Way counterpart (X-ray II), alterations of the cosmic microwave background (CMB) spectrum due to accreting PBHs (including particle dark matter accretion), and FIRAS data. The merger rates of PBH-NS binaries within the context of nDGP models exhibit a marginal increase compared to those predicted by $f(R)$ models. Consequently, when comparing the two theoretical frameworks, it becomes apparent that the nDGP models impose slightly more rigorous constraints on the abundance of PBHs than those inferred from $f(R)$ models.

The expected upper bounds on the PBH fraction from both gravitational models are potentially detectable about {\bf (i)}: $\mathcal{O}(10)$, {\bf (ii)}: $\mathcal{O}(1)$, or {\bf (iii)}: $\mathcal{O}(0.1)$ merger events per year of $(M_{\rm PBH}=5 M_{\odot}, M_{\rm NS}=1.4 M_{\odot})$ in the comoving volume detectable by LVK detectors.
Upon comparison, it is clear that the relevant constraints for case {\bf (i)} in both gravitational models are less stringent than those derived from non-GW data. That is why we have not provided their results in the figure. However, other cases for the upper limit on PBH abundance may remain more stringent than the constraints derived from non-GW data. In this regard, case {\bf (ii)} imposes less stringent constraints than those obtained from X-ray I, X-ray II, and CMB observations (including particle dark matter accretion) for $20\,M_{\odot}\leqslant M_{\rm PBH}\leqslant 100\,M_{\odot}$, and is more stringent than those constraints obtained from non-GW data for $M_{\rm PBH}< 20\,M_{\odot}$. Case {\bf (iii)} is also less stringent than the constraints derived from CMB observations (including particle dark matter accretion) for $50\,M_{\odot}\leqslant M_{\rm PBH}\leqslant 100\,M_{\odot}$, and is more stringent than those constraints obtained from non-GW data for $M_{\rm PBH}< 50\,M_{\odot}$. This suggests that within the parameter space not excluded by non-GW data, it may be feasible to rely on upper limits concerning the expected abundances of PBHs in both gravitational models.
%%%%%%%%%%%%%%%%%%%%%%%%%%%%%%%%%%%%%%%%%%%%%%%%
\section{Conclusions} \label{sec:vi}
%%%%%%%%%%%%%%%%%%%%%%%%%%%%%%%%%%%%%%%%%%%%%%%%
In this study, we have elucidated the merger rate of PBH-NS binaries within the framework of two extensively-studied MG models: Hu-Sawicki $f(R)$ gravity and the nDGP gravity. By employing the key parameters of dark matter halos within MG models, including the halo density profile, halo concentration parameter, and halo mass function, we have demonstrated an enhancement in the PBH-NS merger rate compared to GR. This enhancement stems from an extra scalar degree of freedom in these MG theories, which generates a fifth force that enhances gravitational interactions on cosmological scales.

The results also demonstrate that the merger rate increases as the field strengths, characterized by the MG parameters in the two sets of models, increase. Specifically, this includes the field strength, denoted as $f_{R0}$, for $f(R)$ gravity, and $n=H_0 r_{\rm c}$ for nDGP gravity. Moreover, lower mass PBHs display higher merger rates due to their increased number density for a given mass fraction. Integration the merger rate across the halo weighted by the halo mass function reveals that the total merger rate is significantly higher in MG models compared to GR. Furthermore, an analysis of the redshift evolution suggests that this deviation from GR is more pronounced at higher redshifts.

Upper limits on the abundance of PBHs have been set based on sensitivity bands of LVK detectors. In this context, the predicted merger rates of PBH-NS binaries, as derived from MG theories, impose a constraint on the fraction of PBHs, denoted as $f_{\rm PBH}\gtrsim 0.1$. This implies that the theoretical predictions from MG theories could be useful in determining the abundance of PBHs in the Universe. However, additional studies in multi-messenger astronomy could provide more information on these constraints and help improve our understanding of the complexities involved.

Moreover, compared with the existing non-GW constraints on the abundance of PBHs, our analysis within the MG models could potentially impose more rigorous constraints on the fraction of PBHs with masses below $20\,M_{\odot}$ to be $f_{\rm PBH}\lesssim 0.01$, and for masses below $50\,M_{\odot}$ to be $f_{\rm PBH}\lesssim 0.001$. Additionally, slightly more stringent constraints can be discerned for nDGP models in comparison to $f(R)$ models. These results highlight how MG theories can produce accurate predictions of the merger rate of PBH-NS binaries and the abundance of PBHs. Future observations of GW events could confirm these predictions and elucidate gravitational phenomena beyond GR.

In conclusion, we have demonstrated the impact of large-scale gravitational variations on the merger rate of PBH-NS binaries and constraints on the abundance of PBHs. Despite many uncertainties, the merger rates predicted by MG models emphasize the potential of applying this effect to GW observations to assess the validity of these theories against GR. This implies that upcoming research in this area should leverage the supplementary effects of MG theories to expand the theoretical foundations. However, it should be considered that effective impacts may extend beyond the factors contemplated in this study, and the degeneracy in the parameter space should be considered as a potential uncertainty.
%%%%%%%%%%%%%%%%%%%%%%%%%%%%%%%%%%%%%%%%%%%%%%%%%%%%%%%%%%
\section*{Acknowledgements}
The authors thank the Research Council of Shahid Beheshti University.
%%%%%%%%%%%%%%%%%%%%%%%%%%%%%%%%%%%%%%%%%%%%%%%%%%
%%%%%%%%%%%%%%%%%%%%%%%%%%%%%%%%%%%%%%%%%%%%%%%%%%%%%%%%%%

\bibliography{sample631}{}
\bibliographystyle{aasjournal}

\end{document}